\documentclass[preprint,12pt]{elsarticle}

\usepackage{amssymb}
\usepackage{amsmath}
\usepackage{natbib}
\setcitestyle{authoryear,open={(},close={)}} %Citation-related commands
\usepackage{hyperref}

\journal{Encyclopedia for Condensed Matter Physics}

\begin{document}

\begin{frontmatter}

%% Title, authors and addresses

%% use the tnoteref command within \title for footnotes;
%% use the tnotetext command for theassociated footnote;
%% use the fnref command within \author or \affiliation for footnotes;
%% use the fntext command for theassociated footnote;
%% use the corref command within \author for corresponding author footnotes;
%% use the cortext command for theassociated footnote;
%% use the ead command for the email address,
%% and the form \ead[url] for the home page:
%% \title{Title\tnoteref{label1}}
%% \tnotetext[label1]{}
%% \author{Name\corref{cor1}\fnref{label2}}
%% \ead{email address}
%% \ead[url]{home page}
%% \fntext[label2]{}
%% \cortext[cor1]{}
%% \affiliation{organization={},
%%            addressline={}, 
%%            city={},
%%            postcode={}, 
%%            state={},
%%            country={}}
%% \fntext[label3]{}

\title{Numerical methods for localization}

%% use optional labels to link authors explicitly to addresses:
%% \author[label1,label2]{}
%% \affiliation[label1]{organization={},
%%             addressline={},
%%             city={},
%%             postcode={},
%%             state={},
%%             country={}}
%%
%% \affiliation[label2]{organization={},
%%             addressline={},
%%             city={},
%%             postcode={},
%%             state={},
%%             country={}}

\author{Rudolf A.\ R\"{o}mer}

\affiliation{organization={Department of Physics, University of Warwick},%Department and Organization
            addressline={Gibbet Hill Road}, 
            city={Coventry},
            postcode={CV4 7AL}, 
            state={West Midlands},
            country={United Kingdom}}

\begin{abstract}
%% Text of abstract
Anderson localization provides a challenge to numerical approaches due to the inherent randomness, and hence absence of simple symmetries, in its discrete Hamiltonian representation. Numerous algorithmic approaches have been developed or adopted from other fields and have been collected in this encyclopedia entry. In the discussions below, the emphasis is on the numerical algorithms for localization, while the discussion of the physics of localization is referred to in companion entries by Elgart and Oganesyan in this encyclopedia.
\end{abstract}

%%Graphical abstract
% \begin{graphicalabstract}
% %\includegraphics{grabs}
% \end{graphicalabstract}

%Research highlights
\begin{highlights}
\item the article reviews existing numerical algorithms used when studying Anderson localization and gives relevant references
\item algorithms for data generation
\begin{enumerate}
    \item diagonalization strategies such as exact methods, sparse-matrix methods, kernel-polynomial methods
    \item iterative/quasi-1D algorithms such as the transfer-matrix and the Green function method
    \item renormalization-inspired numerical approaches
\end{enumerate}
\item data analysis algorithms
\begin{enumerate}
    \item energy-level and -ratio statistics
    \item wave function statistics and multi-fractal analysis
    \item finite-size scaling algorithms, with and without assuming a scaling form
\end{enumerate}
\item current challenges
\begin{enumerate}
    \item numerical algorithms in use for data generation and analysis of interacting disordered systems
    \item algorithms based on machine learning strategies
\end{enumerate}
\end{highlights}

\begin{keyword}
%% keywords here, in the form: keyword \sep keyword
exact diagonalization\sep sparse matrix diagonalization\sep recursive diagonalization methods\sep transfer-matrix method\sep Green function method\sep numerical renormalization group methods\sep decimation method\sep energy-level statistics \sep wave function statistics\sep energy ratio statistics\sep  multi-fractal analysis\sep finite-size scaling\sep density-matrix renormalization group\sep strong disorder renormalization group\sep machine learning
%% PACS codes here, in the form: \PACS code \sep code

%% MSC codes here, in the form: \MSC code \sep code
%% or \MSC[2008] code \sep code (2000 is the default)

\end{keyword}

\end{frontmatter}

%% \linenumbers

%% main text

%%%%%%%%%%%%%%%%%%%%%%%%%%%%%%%%%%%%%%%%%%%%%%%%%%%%%%%%%%%%%%%%%%%%%%%%%%%%%%%%%%%
\section{\label{sec-intro}Introduction}

%- hamiltonian \cite{And58}
%- schroedinger equation

Anderson localization, as reviewed in this encyclopedia in the entries by Elgart and Ogenesyan, cannot be solved analytically in dimensions two and above \citep{Stollmann2001CaughtDisorder}. Hence much effort has been focused on developing efficient numerical tools. In its usual form, the discrete Anderson Hamiltonian \citep{And58} is given by the matrix elements
\begin{equation}
    H_{\mathbf{a}\mathbf{b}} = \varepsilon_\mathbf{a} \delta_{\mathbf{a}\mathbf{b}} - t_{\mathbf{a}\mathbf{b}} \delta_{\langle \mathbf{a}, \mathbf{b} \rangle}, 
    \label{eq-hamiltonian}
\end{equation}
where $\varepsilon_\mathbf{a}$ is the potential onsite disorder and $\delta_{\mathbf{a}\mathbf{b}}$ the Kronecker delta, while $t_{\mathbf{a}\mathbf{b}}$ denotes the kinetic hopping energy with $\delta_{\langle \mathbf{a}, \mathbf{b} \rangle}$ non-zero when lattice sites $\mathbf{a}= (a_1, a_2, \ldots, a_d)$, $\mathbf{b}= (b_1, \ldots, b_d)$ are within a defined distance relationship on a discrete, $d$-dimensional lattice $\mathcal{L}_d$ --- usually the nearest-neighbor separation on $\mathcal{L}_d$. With $|\mathcal{L}_d|$ counting the number of lattice sites, $H_{\mathbf{a}\mathbf{b}}$ is an $|\mathcal{L}_d| \times |\mathcal{L}_d|$ matrix.
The time-independent single-particle Schr\"{o}\-din\-ger equation can then be written as
\begin{equation}
    \varepsilon_\mathbf{a} \Psi_\mathbf{a}
    - \sum_{\mathbf{b} \in \mathcal{L}}
    t_{\mathbf{a}\mathbf{b}} \delta_{\langle \mathbf{a}, \mathbf{b} \rangle} \Psi_\mathbf{b}
    = E \Psi_\mathbf{a}
    \label{eq-schrodinger}
\end{equation}
with $\Psi_\mathbf{a} = \Psi_{a_1, \ldots, a_d}$ the coefficients of the wave function $\Psi= \sum_{\mathbf{a}\in \mathcal{L}_d} \Psi_\mathbf{a} | \mathbf{a} \rangle$  in a suitable basis $| \mathbf{a} \rangle$ and $E$ the energy.

\begin{figure}
    \centering
    (a)\includegraphics[width=0.28\columnwidth]{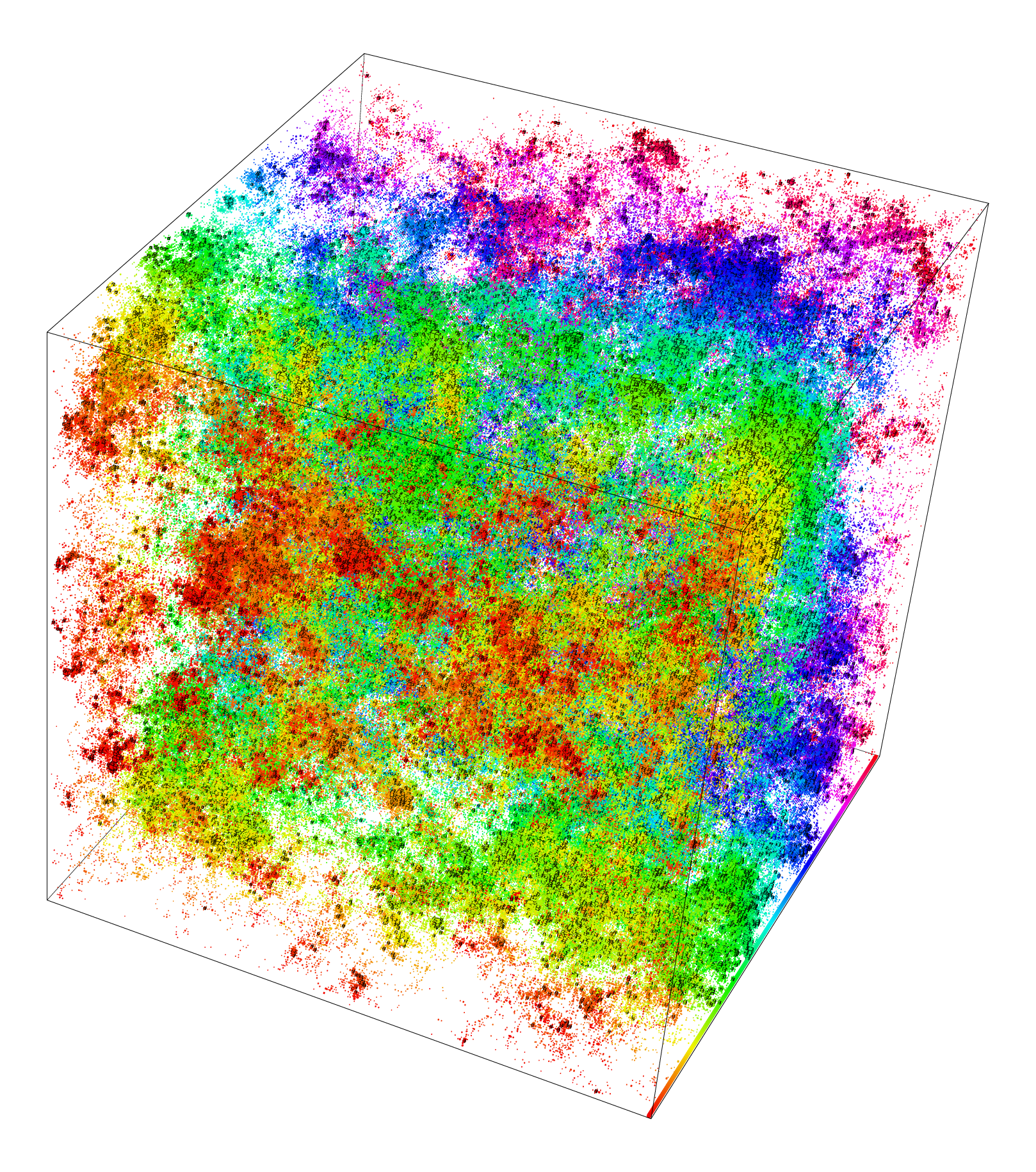}
    (b)\includegraphics[width=0.28\columnwidth]{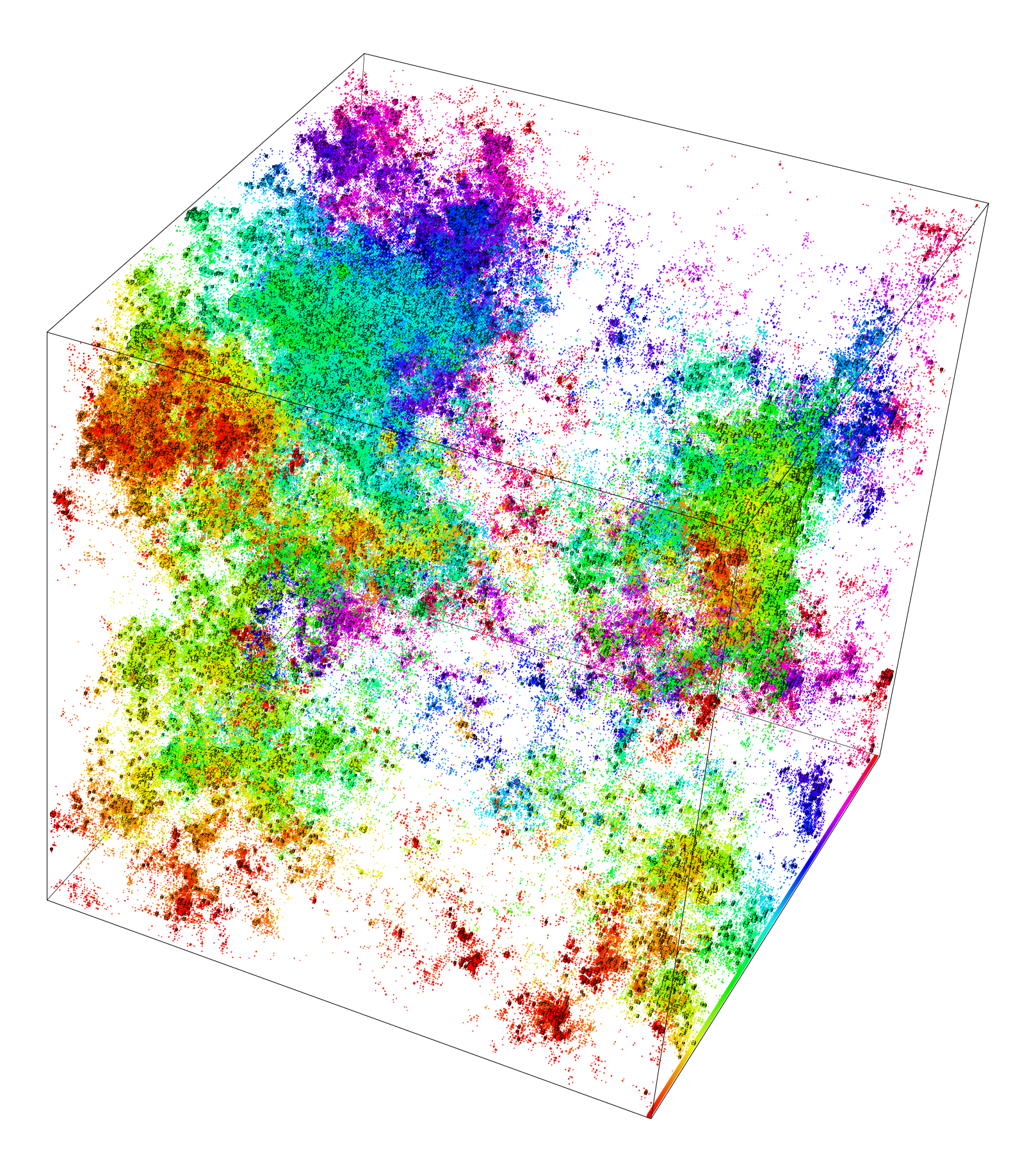}
    (c)\includegraphics[width=0.28\columnwidth]{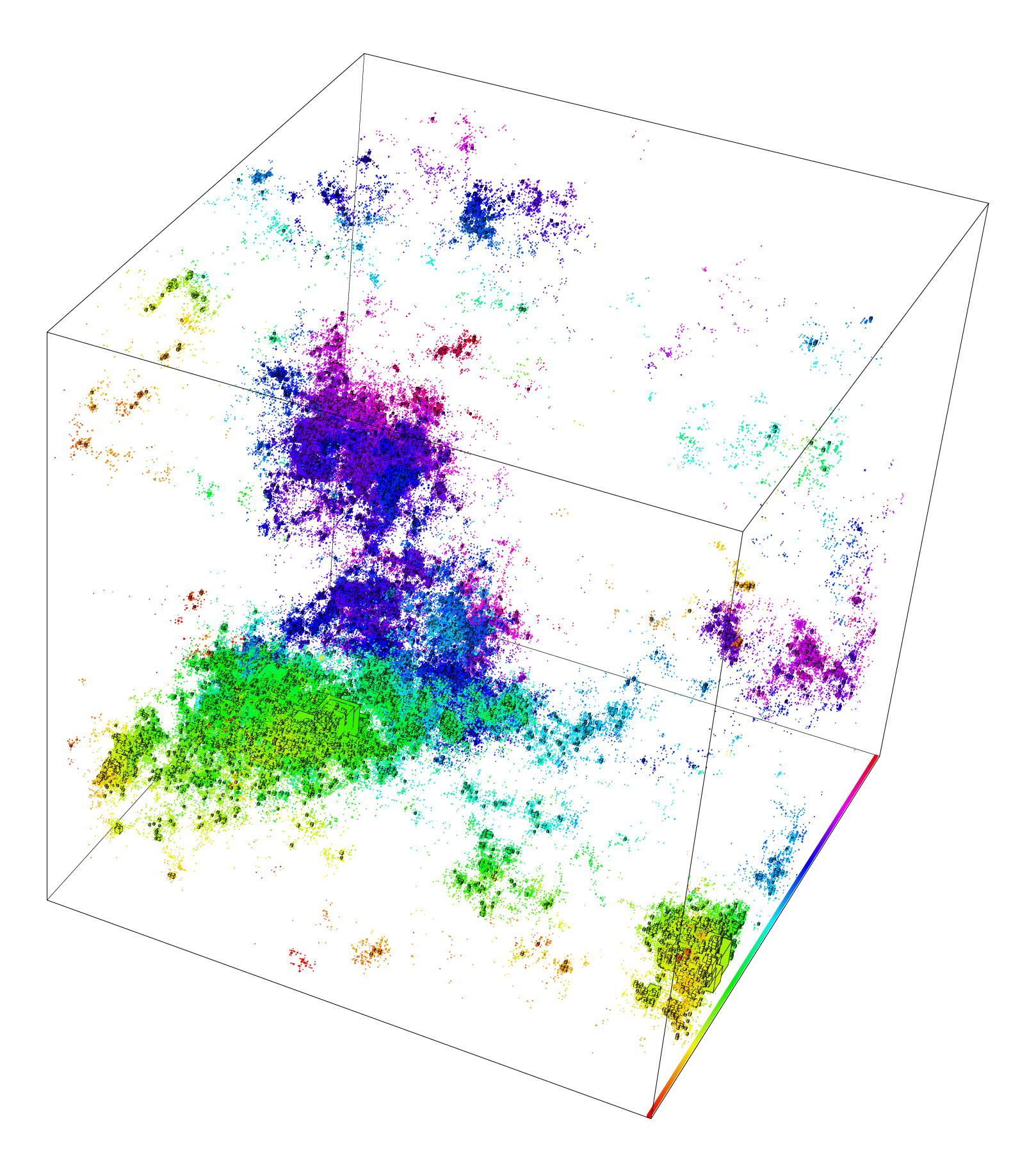}
    \caption{(a) Extended, (b) critical and (c) localized wave function probabilities for the three-dimensional Anderson model with periodic boundary conditions at $E\approx 0$ with $N =200^3$ and $W =15$, $16.5$ and $18$, respectively. Every site $\mathbf{a}$ with probability $|\psi_\mathbf{a}|^2$ larger than the average $1/|\mathcal{L}_3|$ is shown as a box with volume $|\psi_\mathbf{a}|^2 |\mathcal{L}_3|$. Boxes with $|\psi_\mathbf{a}|^2 |\mathcal{L}_3| > \sqrt{1000}$ are plotted with black edges. The color scale along the bottom right edge distinguishes between different slices of the system. The eigenstates have been computed with {\sc Jadamilu} \citep{Bollhofer2007a} as described by \cite{Schenk2008b}.}
    \label{fig:diagonalization}
\end{figure}

In its simplest form, the Hamiltonian \eqref{eq-hamiltonian} is studied on a $d$-cubic lattice with $t_{\mathbf{a}\mathbf{b}}=t_{\mathbf{b}\mathbf{a}}=t=1$ for all nearest-neighbors $\mathbf{a}$, $\mathbf{b}$ and $\varepsilon_\mathbf{a}$ independent and identically distributed random numbers  $\in \left[ -W/2, W/2 \right]$. Hence the parameter $W/t$ expresses the strength of the disorder. See Fig.\ \ref{fig:diagonalization} for examples of $\Psi$ in $d=3$.
Since numerical studies of Anderson localization began to emerge in the late 1970s, this set-up has of course been much varied to include more complicated situations \citep{Krameri1993}. Examples include variations in $d$ from $1$ to $\infty$, changes in the lattice structure, symmetry and topology, changes to and additional disorder or correlations in the $\varepsilon_\mathbf{a}$ and $t_{\mathbf{a}\mathbf{b}}$, as well as inclusion of spin-, external field and interaction effects \citep{Brandes2003AndersonRamifications}. The common theme unifying all the studies is the absence of symmetries in $H$ due to the disorder which in turn prevents a simple restructuring of the matrix into much smaller irreducible blocks \citep{Evers2008}. Connecting the properties of such a single large-block matrix with the physics of Anderson localization is the challenge taken on by the numerical algorithms listed in this encyclopedia entry.

%%%%%%%%%%%%%%%%%%%%%%%%%%%%%%%%%%%%%%%%%%%%%%%%%%%%%%%%%%%%%%%%%%%%%%%%%%%%%%%%%%%
\section{\label{sec-diagonalization}Exact diagonalization}

% - standard
% - sparse (Lanczos \cite{Cullum2002LanczosComputations,DayalTheLibrary}, JADAMILU \cite{Schenk2008b}, FEAST)
% - kernel polynomial \cite{Weie2006}

Given an invertible matrix \eqref{eq-hamiltonian}, the most straightforward approach is to use standard \emph{full} matrix methods to compute eigenvalues and eigenvectors such as given by the {\sc LaPack} library routines \citep{Anderson1999LAPACKGuide} and their many highly optimized implementations. At the time of writing this entry, matrices of sizes up to $\sim 10^4 \times 10^4$ can be readily studied with this method. However, as an average over many different disorder realizations is always necessary to find physically reliable results, already such matrix sizes can lead to many weeks of computing. Many of the algorithms mentioned below were conceived to allow for faster availability of results, while at the same time leading to a more limited set of computed information.

When $\delta_{\langle \mathbf{a}, \mathbf{b} \rangle}$ is sufficiently short-ranged, many of the matrix elements  $H_{\mathbf{a}\mathbf{b}}$ are equal to $0$. In this case, it can be advantageous to use \emph{sparse} matrix methods. Such methods construct selected regions of the eigenspectra using repeats of an optimized matrix-vector product computation for $H_{\mathbf{a}\mathbf{b}} \Psi_\mathbf{b}$ following the classic power series algorithm \citep{Horn1985MatrixEdtion}. For the Anderson problem, the implementation of the Lanczos variant by \cite{Cullum2002LanczosComputations} is particularly useful since the disorder removes spectral degeneracies such that the "ghost eigenvalues" coming from the implementation can be readily identified. Coupling the sparse-matrix approach with advantages in iteratively solving coupled systems of equations, \cite{Schenk2008b} found that a Jacobi-Davidson approach can construct eigenstates of size up to $350^3 \times 350^3$ in the numerically most demanding region around the centre of the spectrum within a reasonable time. \cite{Weie2006} showed that similar sizes can be reached using kernel-polynomial methods when coupled with efficient recursive polynomial expansion techniques. The method is particularly efficient in determining a change in the \emph{distribution} of the local density of states as a criterion for Anderson localization.

%%%%%%%%%%%%%%%%%%%%%%%%%%%%%%%%%%%%%%%%%%%%%%%%%%%%%%%%%%%%%%%%%%%%%%%%%%%%%%%%%%%
\section{\label{sec-tmm}Quasi 1D methods}

\begin{figure}[tb]
    \centering
    \includegraphics[width=0.6\columnwidth,angle=90,trim=7cm 5cm 5cm 8cm,clip]{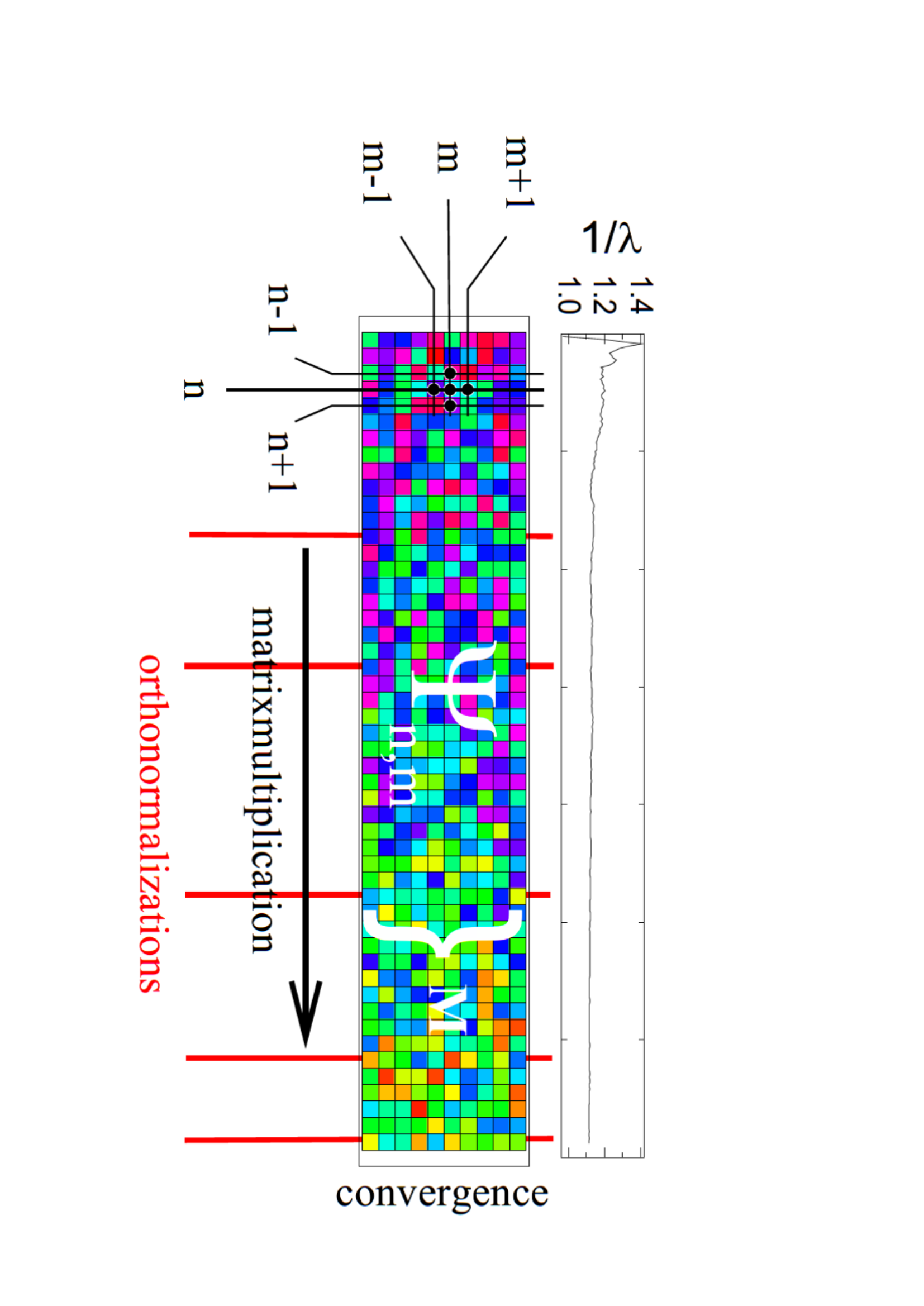}
    \caption{Schematic diagram of the TMM method for the example of a $d=2$ Anderson model. At slice $n$ and position $m$, one needs to know the (past) $\psi_{n-1,m}$ value, the (present) values of $\psi_{n,m}$ and $\psi_{n,m\pm 1}$ and and then compute the (future) value of $\psi_{n+1,m}$. The red lines indicate the orthogonalizations for $\psi$ needed for numerical stability. The diagram at the top shows the behavior of the typical convergence for a Lyapunov exponent measured after each reorthonormalization, the actual number of reorthonormalization used here exceeds $10^4$. The colors of $\psi_{n,m}$ are chosen such that they indicate the eventual convergence for $n\rightarrow \infty$.}
    \label{fig:quasi1d}
\end{figure}

% - Derrida \cite{Derrida1985ProductsSystems} +Pichard \cite{Pichard1981FiniteLocalisation}
% - MacKinnon+Kramer
% - non-square-cubic lattices \cite{Schreiber1992,EILMES2008}
% - diagonal propagation directions \cite{Frahm1995a}
% - BackwardForward method \cite{Ndawana005}
% - Slevin parallel implementation \cite{Slevin2018a}

Repeated matrix-vector computations also underlie the celebrated transfer-matrix method (TMM) \citep{Pichard1981FiniteLocalisation, MacKinnon1983}. The main idea is that it is often possible for a suitable $\mathcal{L}$ to rewrite \eqref{eq-schrodinger} as
\begin{equation}
\left(\begin{array}{l}
    \psi_{n+1} \\
    \psi_{n}
\end{array}\right) = 
\left(\begin{array}{cc}
    \tau_{\langle n+1,n \rangle}^{-1} (E \mathbf{1}-\mathbf{H_{d-1}}) & -\tau_{\langle n+1,n \rangle}^{-1} \tau_{\langle n,n-1 \rangle} \\
    \mathbf{1}             & \mathbf{0}
\end{array}\right)
\left(\begin{array}{l}
    \psi_{n} \\
    \psi_{n-1}
\end{array}\right) .
\label{eq-tmm}
\end{equation}
Here, $\psi_n= (\psi_{n,1}, \ldots, \psi_{n,|\mathcal{L}_{d-1}|})$ are transfer states along a single spatial coordinate in $\mathcal{L}$ while $\mathbf{H_{d-1}}$ is the $|\mathcal{L}_{d-1}| \times |\mathcal{L}_{d-1}|$ Hamiltonian along the remaining $d-1$ directions and $\tau_{\langle n+1, n \rangle}$ denotes the connectivity matrices along the transfer direction (cp.\ Fig.\ \ref{fig:quasi1d}).
The advantage of singling out a special direction along which to study wave function progression lies in the fact that 1D localization is mathematically known to be very strong which usually translates into fast algorithmic convergence.

With $T_n$ the $2|\mathcal{L}_{d-1}| \times 2|\mathcal{L}_{d-1}|$ (local transfer) matrix in \eqref{eq-tmm}, we define the global transfer matrix $\mathcal{T}_{\mathcal{L}_{d-1}}(|\mathcal{L}_1|)= \prod_{n=1}^{|\mathcal{L}_1|} T_n$. Then $\lim_{|\mathcal{L}_1|\rightarrow\infty}(\mathcal{T}\mathcal{T}^\dagger)= \exp[ 2\, \text{diag}(\gamma_1, \ldots,  \gamma_{2\mathcal{L}_{d-1}}) |\mathcal{L}_1|]$ with Lyapunov decay exponents 
$\gamma_1$, 
%\gamma_2$ $\ldots$ $>\gamma_{|\mathcal{L}_{d-1}|}$ $>\gamma_{|\mathcal{L}_{d-1}|+1}$, 
$\ldots$, $\gamma_{2 |\mathcal{L}_{d-1}|}$, ordered 
$\gamma_1 >$ $\gamma_2$ $> \ldots$ $>\gamma_{|\mathcal{L}_{d-1}|}$ $> 0 >$ $-\gamma_{|\mathcal{L}_{d-1}|}$ $>\ldots$ $>-\gamma_{1}$, i.e.\ the Lyapunov exponents come in pairs $\pm \gamma_m$, $m= 1, \ldots, |\mathcal{L}_{d-1}|$ due to the symplectic structure of the $T_n$. The positive $\gamma_{\mathcal{L}_{d-1}}$ indicates the largest extend of the states along the transfer direction, leading to the definition of the (largest) localization length as $\lambda_{\mathcal{L}_{d-1}}= 1/\gamma_{\mathcal{L}_{d-1}}$. Implicit in the convergent construction of the $\lambda_{m}=1/\gamma_m$ is the self-averaging over the disorder along the transfer direction. An error analysis in the \emph{statistical} changes in the $\lambda_{m}$ has to be made when computing the stopping criterion of the TMM \citep{MacKinnon1983}. The $\mathcal{L}_{d-1}$ dependence of $\lambda_{\mathcal{L}_{d-1}}$ lends itself to finite-size scaling and can hence be used to characterize universal properties of Anderson localization. A direct connection to the typical conductance is also known \cite{Pichard1981FiniteLocalisation}.

Generalizations of the TMM include different lattice structures \citep{Schreiber1992,EILMES2008} as well as non-standard transfer directions \citep{Frahm1995a}. For disorder distributions for whom self-averaging is not applicable, a forward/backward variant of the TMM has been given as well \citep{Frahm1995a,Ndawana005}. A recent reevaluation of the idea of self-averaging has also led to the first efficient implementation of TMM on massively-parallel computing architectures \citep{Slevin2018a}.

% %%%%%%%%%%%%%%%%%%%%%%%%%%%%%%%%%%%%%%%%%%%%%%%%%%%%%%%%%%%%%%%%%%%%%%%%%%%%%%%%%%%
% \section{\label{sec-gfm}Green function methods}

% - MacKinnon \cite{MacKinnon1983}
% - von Oppen \cite{}

Computing matrix elements of the resolvent $(E \mathbf{1}-\mathbf{H_{d-1}})^{-1}$ leads to a conceptually very similar recursion setup compared to \eqref{eq-tmm}. This so-called Green function method \citep{MacKinnon1981One-ParameterSystems} again provides access to localization lengths and conductivity for chosen values of $\mathcal{L}_{d-1}$. An implementation of the method along a non-standard transfer direction has been given by \cite{VonOppen1996Interaction-InducedProperties}.

%%%%%%%%%%%%%%%%%%%%%%%%%%%%%%%%%%%%%%%%%%%%%%%%%%%%%%%%%%%%%%%%%%%%%%%%%%%%%%%%%%%
\section{\label{sec-rg}Renormalization and decimation methods}

\begin{figure}
    \centering
    (a) \includegraphics[width=0.45\columnwidth,clip]{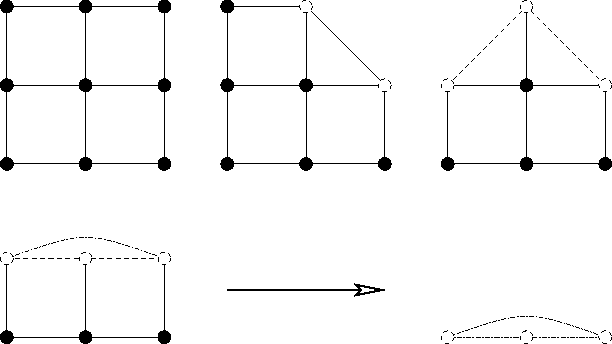} \hfill
    (b) \vspace*{0ex}\includegraphics[width=0.4\columnwidth,clip]{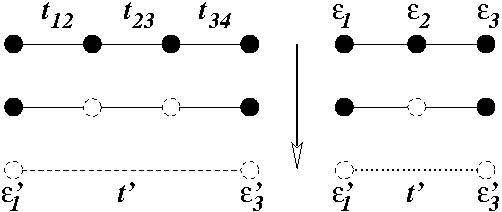}
    \caption{Schematic of real-space decimation procedures using (a) Green function renormalization in $d=2$ and (b) local energy renormalization for hopping (left of $\downarrow$) and onsite terms (right of $\downarrow$) in, e.g., $d=1$. In both schematics, solid lines and circles indicate unrenormalized hopping ($t_{12}$, $t_{23}$, $t_{34}$) and onsite energies ($\varepsilon_1$, $\varepsilon_2$, $\varepsilon_3$), respectively, while the dashed and dotted lines and circles represent renormalized values ($t'$ and $\varepsilon'_1$, $\varepsilon'_3$). The (a) horizontal and (b) vertical arrows represent to renormalization progression in both schemes.}
    \label{fig:renormalization}
\end{figure}

% - Decimation \cite{Leadbeater1999}
% - Brazilian RS RG \cite{Mard2014}
% - CC RS RG \cite{Cain2001b}

Real-space renormalization approaches provide another starting point for studying Anderson localized systems. Here the idea is that an individual lattice site $\mathbf{a}$ can be "removed" from $\mathcal{L}$, leading to a modified $H_{\mathbf{a}\mathbf{b}}$. 
For example, the decimation method \citep{Aoki1982a,Leadbeater1999}, finds
$H'_{\mathbf{a},\mathbf{b}} = H_{\mathbf{a},\mathbf{b}} + \left( H_{\mathbf{a},|\mathcal{L}_d|} H_{|\mathcal{L}_d|,\mathbf{b}} \right) / \left( E  - H_{|\mathcal{L}_d|,|\mathcal{L}_d|} \right)$ after removal of the site $|\mathcal{L}_d|$ while the Green function remains unchanged. In this way, the decimation of lattice sites can proceed until a sufficiently small Hamiltonian has been constructed to use the exact diagonalization routines reviewed above (cp.\ Fig.\ \ref{fig:renormalization}(a)).
Another recent variant of real-space normalization makes use of the information encoded in the $\varepsilon_\mathbf{a}$ and $t_{\mathbf{a},\mathbf{b}}$ values. By recursively elimination those sites with the largest local energy scale $\Omega=\text{max}_{\mathbf{a},\mathbf{b}} \left\{ |\varepsilon_\mathbf{a}|, |t_{\mathbf{a},\mathbf{b}}| \right\}$, \citep{JavanMard2014Strong-disorderModel,Mard2014} continue until only a single renormalized link $\varepsilon'_{\alpha}$ -- $t'_{\alpha,\beta}$ -- $\varepsilon'_{\beta}$ is left. This last link can then be used to compute transport properties, including their disorder $W$ dependence and the scaling behaviour (cp.\ Fig.\ \ref{fig:renormalization}(b)). The method is applicable for $1 \leq d < \infty$. In $d=2$, the approach is similar in spirit to the real-space renormalization approach used for localization in a magnetic field \citep{Cain2005}.

%%%%%%%%%%%%%%%%%%%%%%%%%%%%%%%%%%%%%%%%%%%%%%%%%%%%%%%%%%%%%%%%%%%%%%%%%%%%%%%%%%%
\section{\label{sec-els}Energy-level statistics}

\begin{figure}
    \centering
    (a)\includegraphics[width=0.45\columnwidth]{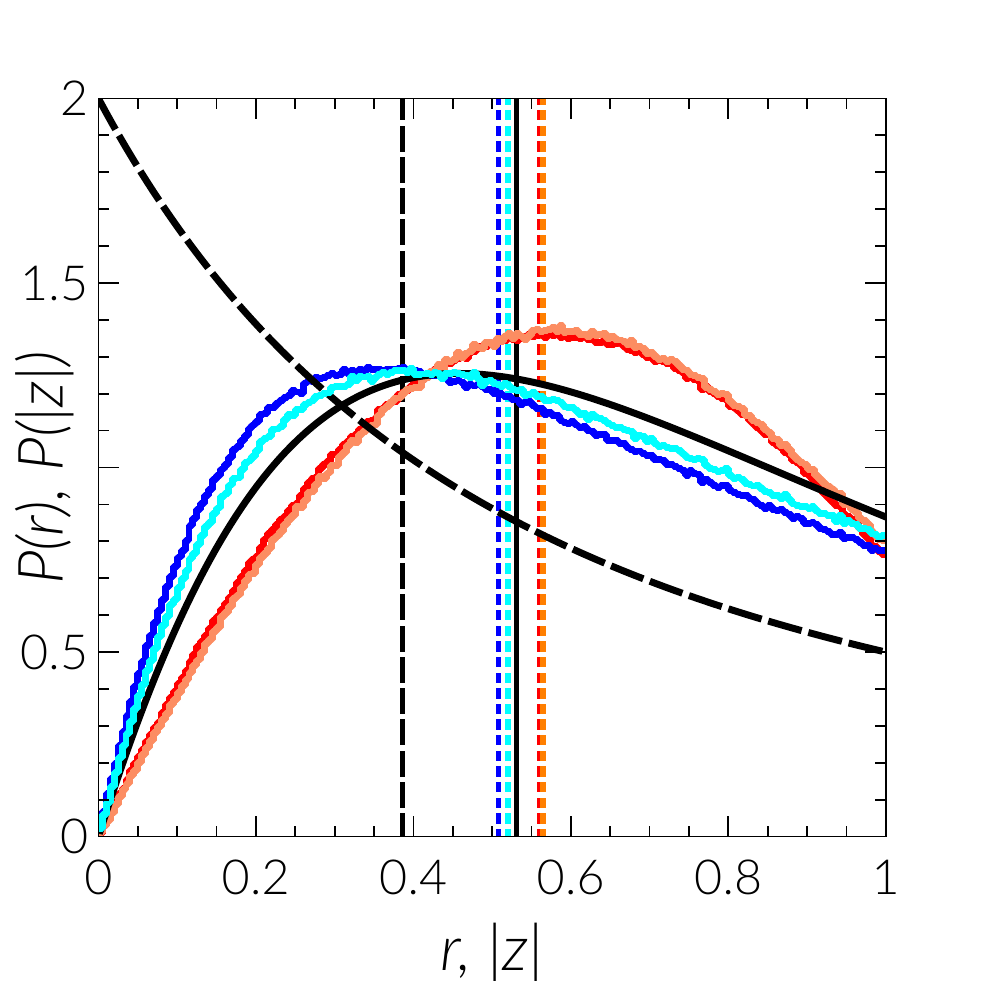}
    (b)\includegraphics[width=0.45\columnwidth]{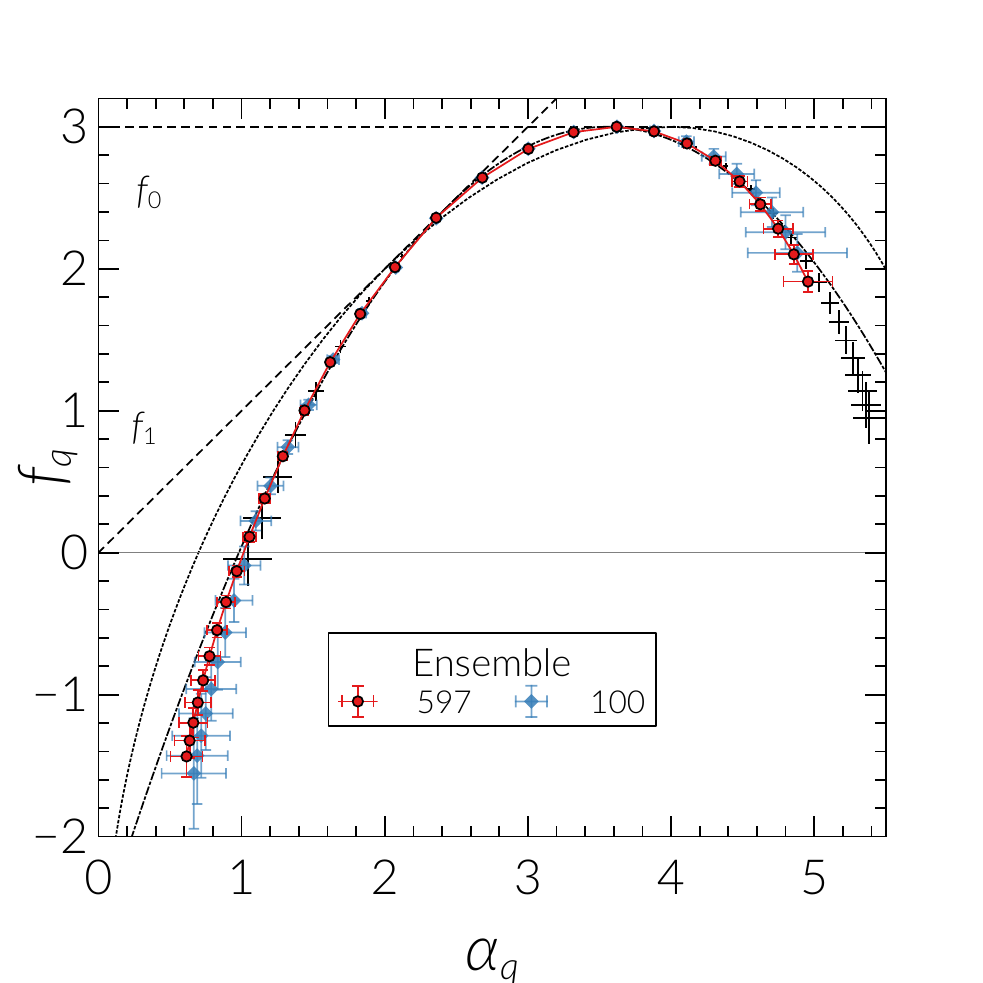}
    \caption{
    (a) Distributions of the energy level ratios $r$ (dark and light blue) and $|z|$ (dark and light red) for a $d=3$ Lieb model with $44^3$ sites \citep{Liu2020} and averaged over $\sim 8\times 10^6$ ratios for each curve. Darker colors denote slightly larger disorders. 
    The black dashed and solid lines denote the exact $P_\text{Poisson}(r)$ and a surmise for $P_\text{GOE}(r)$. No such predictions are yet known for $P(|z|)$ \citep{Luo2021}.
    (b) Ensemble-averaged singularity spectrum $f_q(\alpha_q)$, parameterised by $q$-moments, for Kohn-Sham wave functions computed via density-functional theory of Si:P with $22^3$ atoms of which $140$ are P impurities \citep{Carnio2018}.
    The spectrum has been sampled for values of $q=5, 4.75, \ldots, -1.75, -2$ (decreasing from left to right) in the bulk of the impurity band at $0.2495$ eV below the Fermi energy.
    Blue diamonds show the results for the ensemble of the first $100$ disorder realisations, while red circles indicate the results from all $597$ available realisations. 
    Simple error bars, without data point, indicate the symmetrised spectrum of the full ensemble. Dashed lines indicate the functions $f_0\equiv 3$ and $f_1(\alpha)= \alpha$. 
    The dotted line represents the spectrum for the standard cubic 3D Anderson model at criticality, reproduced from \cite{Rodriguez2011}, while the dot-dashed line shows the fit to a parabolic approximation \citep{Evers2008}. 
    The two horizontal lines are guides to $f=0$ and $3$.
    }
    \label{fig:energylevelstatistics}
    \label{fig:wavefunctionstatistics}
\end{figure}

% - P(s) \cite{Mehta2004RandomMatrices}
% - unfolding
% - P(s) stats, Delta2, Sigma2 \cite{Mehta2004RandomMatrices}
% - P(r) (and advantage of using sparse) \cite{Oganesyan2007LocalizationTemperature}
% - P(q) (and advantage of using sparse) \cite{Luo2021UniversalityDisorder}

The distribution of energy-level spacings $\delta_n = E_{n+1} - E_{n}$ in the eigenspectra $\{ E_n \}$ of Anderson-type systems has long been known to provide a marker able to distinguish between extended and localized eigenstates $\Psi$ \citep{Altshuler1986RepulsionSamples,Evangelou1992SpectralEnsemble,Shklovskii1993StatisticsTransition} with distribution functions $P(\delta)$ of the spacing following the three Dyson ensembles --- e.g.\ with the Gaussian orthogonal ensemble (GOE) corresponding to real-symmetric matrices invariant under orthogonal transformations --- \citep{Dyson1962StatisticalI,Shklovskii1993StatisticsTransition} and generalizations thereof \citep{Altland1997NonstandardStructures}. Often, the integrated distribution function $I(\delta)= \int_{0}^{\delta} P(\delta') d\delta'$ is a numerically more stable indicator \citep{ZHONG1998}. More detailed tests of the spectral properties such as the $\Delta_3$ and $\Sigma_2$ statistics have also been employed \citep{Mehta2004RandomMatrices,Hofstetter1993StatisticalHamiltonian}. As usual, the $|\mathcal{L}_d|$ dependence of these quantities can be used for scaling; convenient estimates of $P(s)$, etc., have been given based on Wigner surmises \citep{Shklovskii1993StatisticsTransition}.

However, before a detailed comparison with predictions of random matrix theory is possible, the region of the spectrum under investigation has to be \emph{unfolded} so that the unfolded density-of-states is constant \cite{Hofstetter1993StatisticalHamiltonian}. This adds a possible source of numerical uncertainty for systems with large fluctuations in the density-of-states. Recently, it has been suggested to study ratios of level spacings such as $0 \leq r_n= \text{min} \left\{ \delta_n, \delta_{n-1} \right\} / \text{max} \left\{ \delta_n, \delta_{n-1} \right\} \leq 1$ \citep{Oganesyan2007b} and others \citep{Luo2021} and their distributions instead (cp.\ Fig.\ \ref{fig:energylevelstatistics}(a)). For such ratios, the unfolding procedure is no longer necessary, while mean and moments are still accessible via surmises and even exact results exist \citep{Atas2013JointMatrices,Giraud2022ProbingStatistics}.

%%%%%%%%%%%%%%%%%%%%%%%%%%%%%%%%%%%%%%%%%%%%%%%%%%%%%%%%%%%%%%%%%%%%%%%%%%%%%%%%%%%
\section{\label{sec-mfa}Wavefunction statistics and multi-fractal analysis}

% \begin{figure}
%     \centering
%     \includegraphics{mfa-fig.pdf}
%     \caption{Ensemble-averaged singularity spectrum $f_q(\alpha_q)$, parameterised by $q$-moments, for Kohn-Sham wave functions computed via density-functional theory of Si:P with $22^3$ atoms of which $140$ are P impurities \citep{Carnio2018}. The spectrum has been sampled for values of $q= 5, 4.75, \ldots, -1.75, -2$ (decreasing from left to right) at energy $−0.249$ eV. Blue diamonds show the results for the ensemble of the first $100$ disorder realisations, while red circles indicate the results from all $597$ available realisations. Simple error bars, without data point, indicate the symmetrised spectrum of the full ensemble. Dashed lines indicate the functions $f_0\equiv 3$ and $f_1(\alpha)= \alpha$. The dotted line represents the spectrum for the standard cubic 3D Anderson model at criticality, reproduced from \cite{Rodriguez2011}, while the dot-dashed line shows the fit to a parabolic approximation \citep{Evers2008}.}
%     \label{fig:wavefunctionstatistics}
% \end{figure}

% - IPR \cite{Wegner1980}
% - PT \cite{StatisticalSpectra} + 1/g stats \cite{Mirlin1999CorrelationsSystems}
% - MFA \cite{Rodriguez2011}

Unsurprisingly, the properties of the eigenstates of \eqref{eq-hamiltonian} are also useful when studying the localization properties. In the simplest situation, the participation number $\mathcal{P}= \left( \sum_{\mathbf{a}\in \mathcal{L}_d} | \Psi_\mathbf{a} |^4 \right)^{-1}$ \citep{Wegner1980} measures how many sites in $\mathcal{L}_d$ \emph{participate} in the spatial extend of a given $\Psi$. E.g.\ for a localized $\Psi$, $\mathcal{P}\sim 1$, whereas for an extended $\Psi$, $\mathcal{P}\sim |\mathcal{L}_d|$.
The universal statistical properties of $\Psi$ can be computed as well, with the Porter-Thomas distributions capturing the behaviour at weak disorder while the systematic corrections at larger disorder have been calculated in a series of works \citep{Mirlin1999CorrelationsSystems}. 
Directly at the Anderson metal-insulator transition, e.g.\ in three-dimensions, $\Psi$ has the scaling properties of a multifractal \citep{Janssen1998StatisticsSystems}. The full multifractal spectrum can be shown to be system-size independent and the moments of the multifractal distributions can be used to characterize the critical properties of the transition \cite{Rodriguez2011}, via finite-size scaling. A typical multi-fractal $f(\alpha)$ spectrum is shown in Fig.\ \ref{fig:wavefunctionstatistics}(b).

%%%%%%%%%%%%%%%%%%%%%%%%%%%%%%%%%%%%%%%%%%%%%%%%%%%%%%%%%%%%%%%%%%%%%%%%%%%%%%%%%%%
\section{\label{sec-fss}Finite-size scaling}

\begin{figure}
    \centering
    (a)\hspace*{-1ex}\includegraphics[width=0.28\columnwidth]{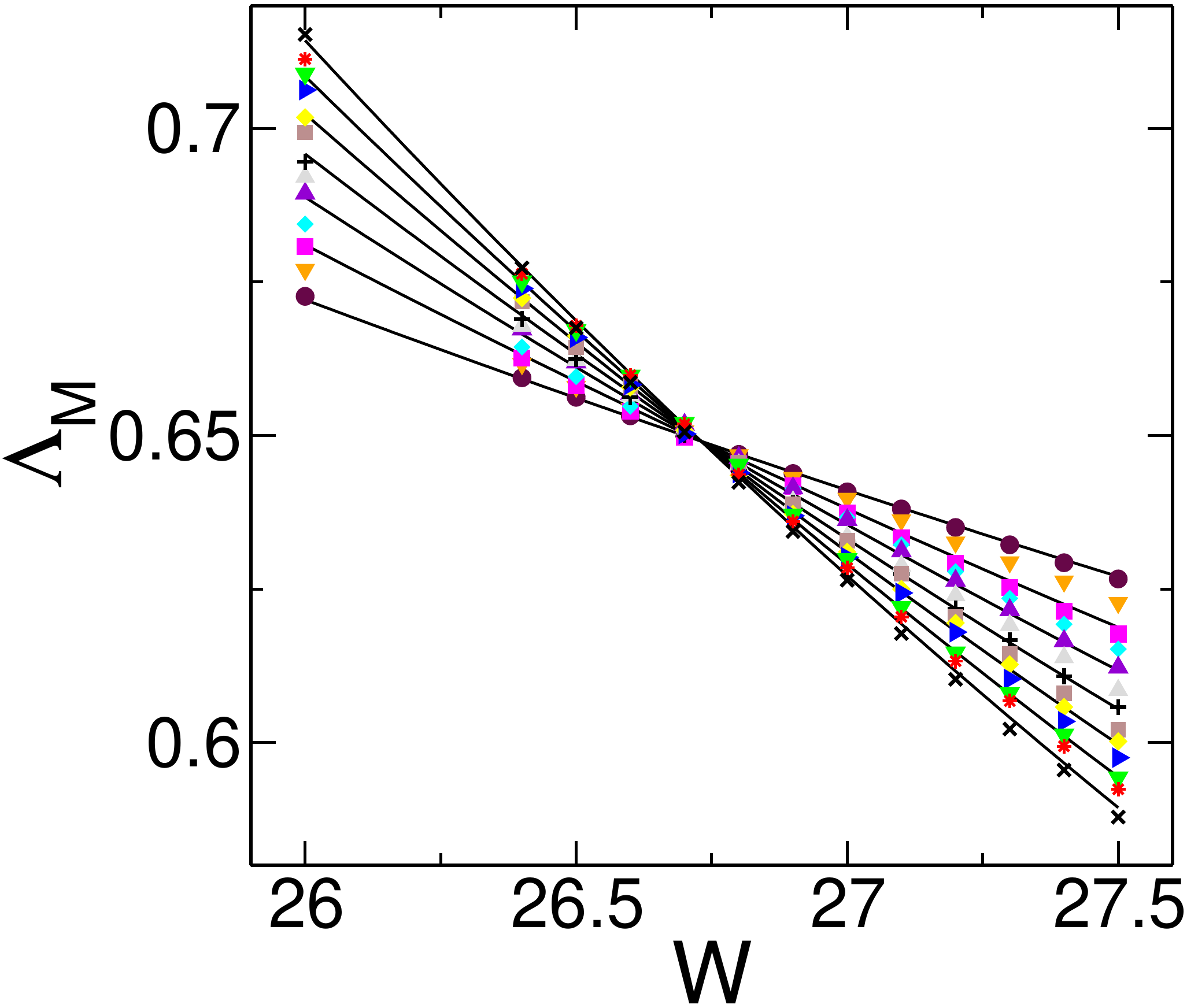} \hfill
    (b)\hspace*{-1ex}\includegraphics[width=0.28\columnwidth]{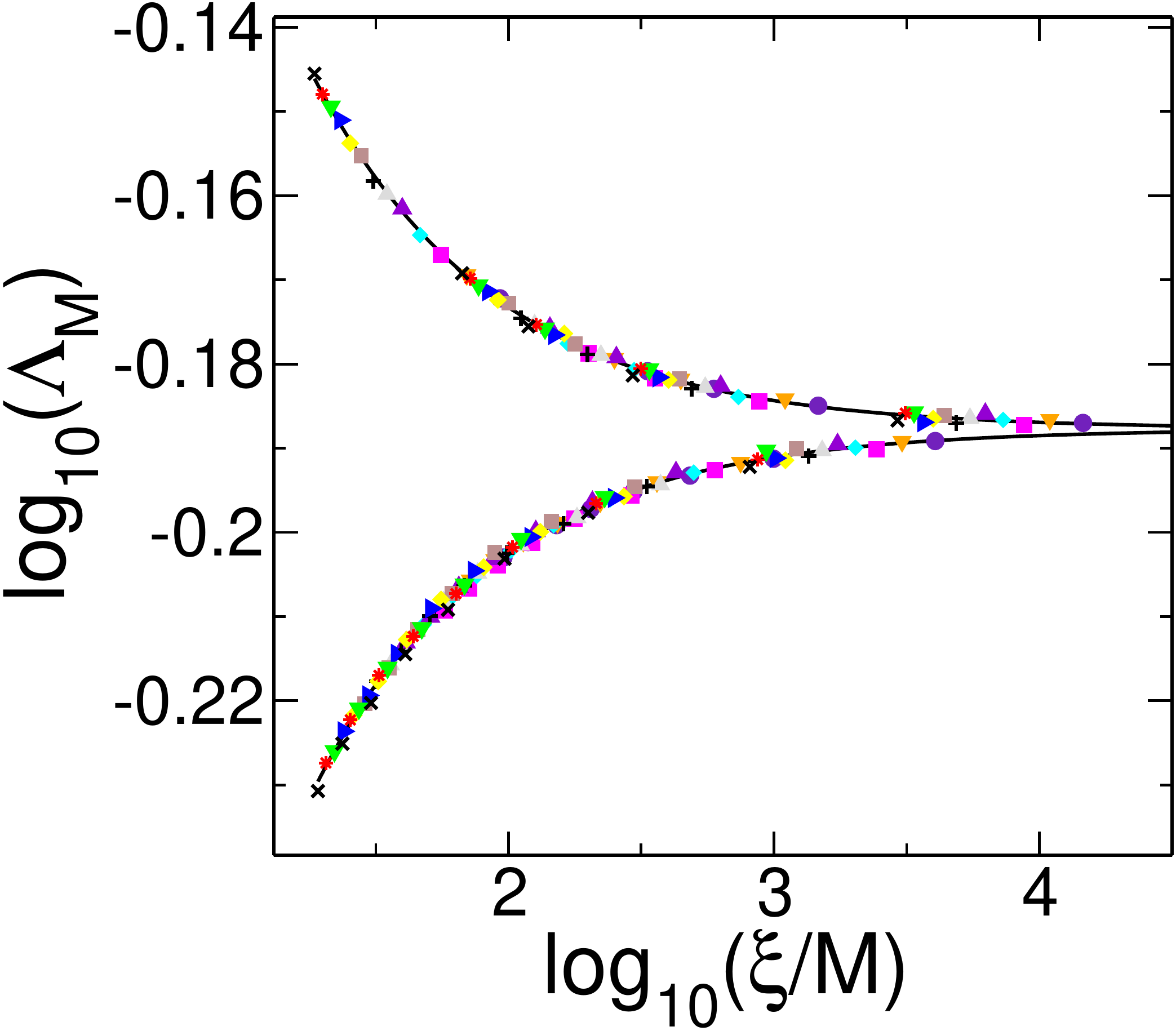} \hfill
    (c)\hspace*{-1ex}\includegraphics[width=0.28\columnwidth]{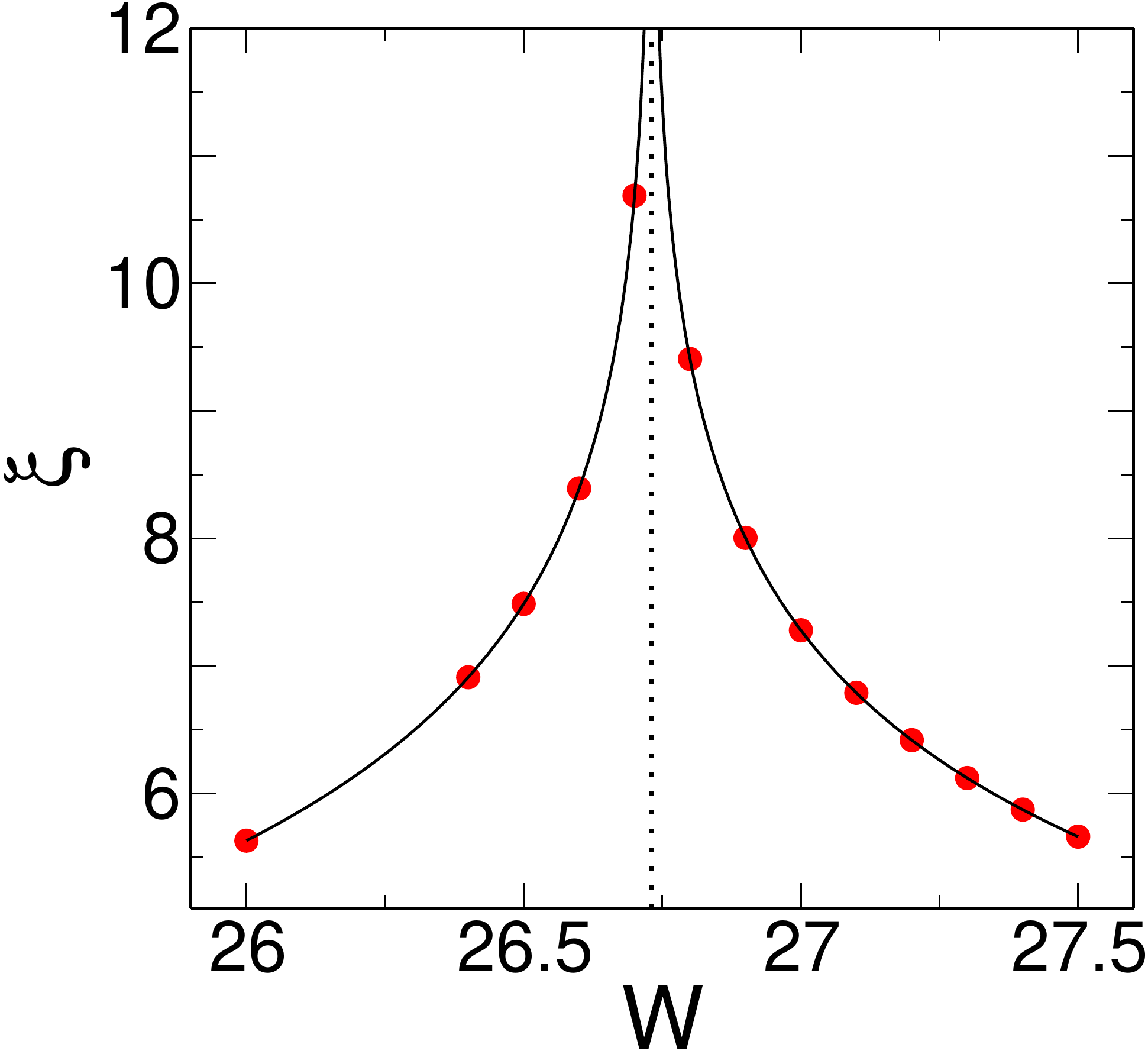}
    \caption{Reduced localization lengths $\Lambda_{M}=\lambda_{|\mathcal{L}_2|}/M$, with $M=|\mathcal{L}_2|$ the size of the TMM bar, versus (a) $W$ or (b) $\xi(W)/M$ for $E=0$ of a $d=3$ Anderson model defined on an fcc lattice \citep{EILMES2008}. Different symbols and colors denote increasing $M= 3 (\bullet), \ldots, 15 (\times)$. Panel (c) shows the constructed $\xi(W)$ scaling parameter function (black line) with divergence at $W_c=26.73(1)$ (vertical dotted line) and the computed $\xi(W_i)$ values for $W_i= 26, \ldots, 27.5$ (circles).}
    \label{fig:fss}
\end{figure}

% - MacKinnon+Kramer \cite{MacKinnon1983}
% - Ohtsuki+Slevin \cite{Slevin1999a}

The finite-size scaling approach to second order phase transitions has been well documented by now \citep{Abrahams1979,Belitz1994a,Slevin1999a}. Roughly speaking, finite-size scaling implies to collapse of data for different parameters such as $E$ and $W$ onto a single scaling curve when their system size $|\mathcal{L}_d|$ dependence is adjusted as well \citep{Slevin1999a}.

For the Anderson localization problem, there exist two different numerical approaches. 
In the absence of a phase transition, i.e.\ in $d \leq 2$, or when the functional form of the transition is not known, one can attempt to simply construct the data collapse of a suitably chosen quantity, such as the localization length $\lambda_{|\mathcal{L}_d|}$ or a multifractal moment $\tau_q$, by minimization of the overlap between curves. In this approach, the shift needed to achieve this minimal overlap defines a scaling parameter $\xi$ and its parameter dependence on $E$ or $W$. For the Anderson transition in three dimensions, the behaviour should be a good approximation of $\xi\propto |E - E_c|^{-\nu}$ or $\xi\propto |W - W_c|^{-\nu}$. This defines the critical transition points $E_c$, $W_c$ as well as the critical exponent $\nu$.

The second, more modern approach, starts by assuming a scaling form for $\xi_{|\mathcal{L}_d|}$, say 
\begin{equation}
    \xi_{|\mathcal{L}_d|}(F)= a_{00}+a_{01} \left(F/F_c -1 \right) {|\mathcal{L}_d|}^{1/\nu} + {|\mathcal{L}_d|}^{-y} \left[ a_{10} + a_{11} \left(F/F_c -1\right) {|\mathcal{L}_d|}^{1/\nu} \right] ,
\end{equation}
and then fits the parameters $a_{ij}$, $\nu$ and $y$ to allow the best possible fit to the accumulated data points by a non-linear regression \citep{Slevin1999a} for either energy $(F=E)$ or disorder ($F=W$) dependence (cp.\ Fig.\ \ref{fig:fss}). This more systematic approach allows the inclusion of the irrelevant scaling exponent $y >0$ as shown in the example. The accuracy of a fit is quantified via $\chi^2$ analysis. When constructing the scaling function $\xi$ in powers of relevant and irrelevant corrections, one has to take care that the fit with the best $\chi^2$ statistics is also robust with respect to changes in the considered range of energies and disorders considered as well as stable within the computed accuracy when increasing the complexity of the fit function. 
It is this second finite-size scaling approach, together with high accuracy data computed by the TMM or sparse matrix diagonalization coupled with multifractal analysis as outlined above which currently gives the accepted best-fit estimates of the critical properties at the Anderson transition \citep{Slevin1999a}.

%%%%%%%%%%%%%%%%%%%%%%%%%%%%%%%%%%%%%%%%%%%%%%%%%%%%%%%%%%%%%%%%%%%%%%%%%%%%%%%%%%%
\section{\label{sec-}Localization and many-body interactions}

% \begin{figure}
%     \centering
%     \includegraphics{placeholder.png}
%     \caption{mlb}
%     \label{fig:mbl}
% \end{figure}

% - ED (MBL)

% - self-consistent Hartree-Fock \cite{Oswald2020,Harashima}
% - DFT \cite{Carnio2017b}

% - DMRG/MPS \cite{White1992DensityGroups,Schollwoeck2004}
% - tSDRG \cite{Goldsborough2014a}

The possibility of including the influence of interactions on the Anderson localization problem has always intrigued the community and was mentioned already by \cite{And58}. Numerically, the problem becomes much harder since the number of states in Hilbert space grows from $\propto |\mathcal{L}_d|^d$ to $\propto \exp |\mathcal{L}_d|$ while the requirement to perform the disorder averaging remains.
Early investigations concentrated on studying the influence of disorder on the ground states of interacting systems. Mostly, progress was based on numerical methods adopted from clean interacting systems with methods such as the density-matrix renormalization group \citep{White1992DensityGroups,Schollwoeck2004} and exact diagonalization most commonly used. For the many-body localization problem (see Oganesyan article in this encyclopedia), the complete spectrum is often of much interest, so the exact diagonalization is often the method of choice, severely restricting the spatial size of systems while still retaining a relatively large Hilbert space \citep{Oganesyan2007b}. Alternative renormalization schemes such as implementations of the strong-disorder renormalization group \citep{Hikihara1999NumericalChains,Goldsborough2014a}, already mentioned above for the non-interacting case, might also provide a way forward.

More recently, approximate methods such as the time-honoured Hartee-Fock \citep{Weidinger2018Self-consistentLocalization,Oswald2020} as well as density-functional-theory methods \citep{Harashima2014,Carnio2017b} have begun to be applied. Particularly the latter offer the exciting possibility of providing a material-specific approach to studies of Anderson localization \citep{Carnio2018}. In this context, one should also mention the typical medium dynamical cluster approximation for disordered electronic systems \citep{Aguiar2008} which can now also be used to incorporate density-functional-theory-derived potentials, the effect of multiple bands, non-local disorder, and electron-electron interactions \citep{Terletska2018}.
Another approximate method that incorporates the full spatial variations of wave functions and \emph{local} interactions effects is the Statistical Dynamical Mean Field Theory \citep{Dobrosavljevic1997MeanTransition,Miranda2012DynamicalDisorder}. Local interactions are included self-consistently through effective correlated single-impurity problems (CSIP) \citep{Georges1996DynamicalDimensions}. In the disordered context, each site defines a different CSIP whose local self-energy is fed back into the lattice self-consistent loop. In the non-interacting limit the formulation is exact. It has been used to study the disordered Mott transition \citep{Miranda2012DynamicalDisorder,Suarez-Villagran2020Two-dimensionalTransition} and disordered heavy fermion systems \citep{Miranda2001Localization-InducedLattices}, where a distribution of Kondo temperatures is a conceptually useful byproduct.

%%%%%%%%%%%%%%%%%%%%%%%%%%%%%%%%%%%%%%%%%%%%%%%%%%%%%%%%%%%%%%%%%%%%%%%%%%%%%%%%%%%
\section{\label{sec-ml}Machine learning}

% \begin{figure}
%     \centering
%     \includegraphics{placeholder.png}
%     \caption{ml}
%     \label{fig:ml}
% \end{figure}

% - supervised for phases \cite{Ohtsuki}
% - unsupervised, etc.

A seemingly new approach has become available to the numerical study of Anderson localization with the emergence of the machine learning paradigm. Instead of specifying in detail which physical aspects of data to concentrate on, say transport properties such as $\lambda$, conductivity or multifractal moments $\tau_q$, the data-driven machine learning approach can be used to find Anderson transitions for variants of the model Hamiltonian \eqref{eq-hamiltonian} when training deep learning networks on, e.g., the standard cubic Anderson model \cite{Ohtsuki2016}. With this form of \emph{supervised} machine learning of Anderson localization, \cite{Ohtsuki2019a} found that one can determine phase diagrams in other disordered systems such as topological insulators and disordered Weyl semimetals (see the entries by ??? in the encyclopedia). How other machine learning strategies such as unsupervised and reinforcement learning might add to the numerical approaches for Anderson localization is currently under much investigation.

%%%%%%%%%%%%%%%%%%%%%%%%%%%%%%%%%%%%%%%%%%%%%%%%%%%%%%%%%%%%%%%%%%%%%%%%%%%%%%%%%%%
\section{\label{sec-conclusions}Conclusions}

In this entry to the encyclopedia, I have attempted to provide a very brief review of the rich variety of numerical algorithms developed to study Anderson localization. Due to brevity requirements appropriate for such an entry, details had to be mostly neglected. Still, with citations provided to hopefully the original publications as well as some recent examples, I hope that the reader can find enough information to get started with each approach and that seeing them in context allows for a considered choice of which method to use.

%%%%%%%%%%%%%%%%%%%%%%%%%%%%%%%%%%%%%%%%%%%%%%%%%%%%%%%%%%%%%%%%%%%%%%%%%%%%%%%%%%%
\subsection*{\label{sec-acknowledgments}Acknowledgments}
I am grateful for constructive suggestions on the manuscript by H.\ Fehske, E.\ Miranda, A.\ Rodriguez and T.\ Ohtsuki. Special thanks to E.\ Carnio for help with Fig.\ \ref{fig:wavefunctionstatistics}.

%% The Appendices part is started with the command \appendix;
%% appendix sections are then done as normal sections
%% \appendix

%% \section{}
%% \label{}

%% If you have bibdatabase file and want bibtex to generate the
%% bibitems, please use
%%
%\bibliographystyle{elsarticle-harv} 
\bibliography{EncCMP-NumLoc}

%% else use the following coding to input the bibitems directly in the
%% TeX file.

% \begin{thebibliography}{00}

% %% \bibitem[Author(year)]{label}
% %% Text of bibliographic item

% \bibitem[ ()]{}

% \end{thebibliography}
\end{document}